\documentstyle[prl,aps]{revtex}
\input BoxedEPS.tex
\SetOzTeXEPSFSpecial
\HideDisplacementBoxes

\begin{document}

\twocolumn[\hsize\textwidth\columnwidth\hsize
           \csname @twocolumnfalse\endcsname
\title{Oxygen isotope effects in high-quality thin films of
manganites: Quantitative constraints on the physics of manganites}
\author{G. M.  Zhao$^{(1,2)}$, D. J. Kang$^{(1)}$, W. Prellier$^{(1,*)}$,
M. Rajeswari$^{(1)}$, }
\author{H. Keller $^{(2)}$, T. Venkatesan $^{(1)}$ and  R. L. Greene$^{(1)}$}
\vspace{1cm}
\address{$^{(1)}$Center for Superconductivity Research, University of
Maryland, College Park, MD 20742,
USA\\
$^{(2)}$ Physik-Institut der Universit\"at Z\"urich, CH-8057
Z\"urich, Switzerland}

\maketitle
\noindent
\begin{abstract}
Oxygen isotope effects on the transport properties have been
studied in high-quality epitaxial thin films of
La$_{0.75}$Ca$_{0.25}$MnO$_{3}$ and Nd$_{0.7}$Sr$_{0.3}$MnO$_{3}$. In
the paramagnetic state, the resistivity can be well fitted by $\rho
(T) = (A/\sqrt{T})\exp(E_{\rho}/k_{B}T)$ with the parameters $A$ and
$E_{a}$ depending strongly on the oxygen isotope mass. The resistivity below
80 K almost perfectly follows $\rho = \rho_{o}+
B\omega_{s}/\sinh^{2}(\hbar\omega_{s}/2k_{B}T)$
with $\hbar\omega_{s}/k_{B}$ $\sim$ 100 K. Both $\rho_{o}$ and $B$
increase by about 15(3)$\%$ upon raplacing $^{16}$O by $^{18}$O. The
results provide quantitative constraints on the basic physics of
manganites.

\end{abstract}
\vspace{1cm}
]

\narrowtext
The discovery of ``colossal" magnetoresistance (CMR) in thin films of
Re$_{1-x}$A$_{x}$MnO$_{3}$ (Re = a rare-earth element, and A = a divalent
element) \cite{Von} has stimulated extensive studies of magnetic,
structural and
transport properties of these
materials \cite{Art}. The physics of manganites
has primarily been described by the
double-exchange (DE) model \cite{Zener}. However, Millis, Littlewood and
Shraiman \cite{Millis1} pointed out that the carrier-spin
interaction in the DE model is too weak to lead to the carrier
localization in the paramagnetic state, and thus a second mechanism such as
small polaronic effects
should be involved to explain the observed resistivity data
in doped manganites. The central point of the model is that in the
paramagnetic
state the electron-phonon coupling constant $\lambda$ is large enough
to form small polarons while the growing ferromagnetic order increases
the bandwidth and thus decreases $\lambda$ sufficently to form
a large polaron metallic state. Many
recent experiments
\cite{Jaime,ZhaoNature,ZhaoPRL,ZhaoJPCM,Billinge,Teresa,Booth} have
provided strong evidence for the
existence of small polarons in the paramagnetic state, and
qualitatively support these theoretical models \cite{Millis1,Roder,Moreo}.

Alexandrov and Bratkovsky \cite{Alex} have recently argued that the
model suggested by Millis {\em et al.,} cannot quantitatively
explain CMR since the characteristic theoretical
field ($\sim$ 15 T) for CMR is too high compared with the experimental
one ($\sim$ 4 T) \cite{Millis1}. They thus proposed an
alternative theory for CMR.
The basic idea of their model is that the small polarons
form localized bound pairs (bipolarons) in the paramagnetic state
due to strong electron-phonon coupling while the competing exchange
interaction
of polaronic carriers with localized spins drives the ferromagnetic
transition. The transition is accompanied by a giant increase in the
number of small polarons which are mobile carriers and move coherently
at low temperatures. This model appears to be able to
explain the CMR quantitatively.

In order to discriminate between these different models, we study the oxygen
isotope effect on the transport properties of high-quality
epitaxial thin films of
La$_{0.75}$Ca$_{0.25}$MnO$_{3}$ and
Nd$_{0.7}$Sr$_{0.3}$MnO$_{3}$. The intrinsic resistivity of
these compounds shows a strong dependence on the oxygen isotope mass
in both paramagnetic and ferromagnetic states. A quantitative data
analysis provides essential constraints on the basic physics
of manganites.
\begin{figure}[htb]
    \ForceWidth{6.6cm}
	\centerline{\BoxedEPSF{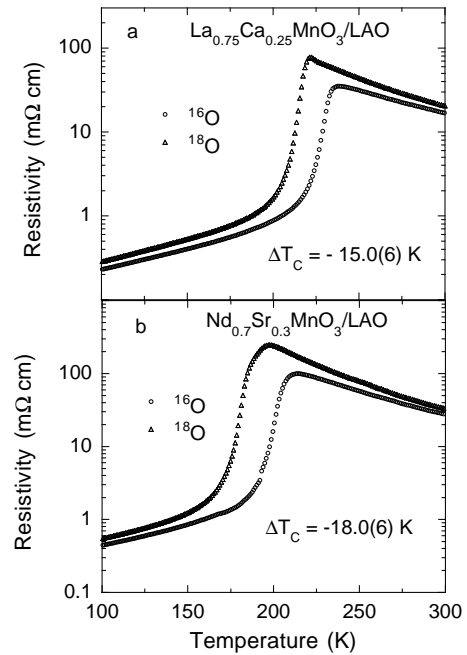}}
	\vspace{0.3cm}
	\caption[~]{The resistivity of the oxygen-isotope exchanged
	films of (a) La$_{0.75}$Ca$_{0.25}$MnO$_{3}$; (b)
Nd$_{0.7}$Sr$_{0.3}$MnO$_{3}$.}
	\protect\label{Fig.1}
\end{figure}

The epitaxial thin films of La$_{0.75}$Ca$_{0.25}$MnO$_{3}$ (LCMO) and
Nd$_{0.7}$Sr$_{0.3}$MnO$_{3}$ (NSMO) were grown
on $<$100$>$ LaAlO$_{3}$ single crystal substrates by pulsed laser
deposition using a KrF excimer laser \cite{Prellier}.
The film thickness was about 190 nm for NSMO and 150 nm for LCMO.
Two halves were cut
from the same piece of a film for oxygen-isotope diffusion.
The diffusion for LCMO/NSMO was
carried out for 10 h
at about 940/900 $^{\circ}$C and oxygen pressure of about 1 bar. The
$^{18}$O-isotope gas is enriched with
95$\%$ $^{18}$O, which can ensure 95$\%$ $^{18}$O in the $^{18}$O thin
films. The resistivity was measured using the van der Pauw technique, and the
contacts were made by silver paste. The measurements were
carried out from 5 to 380 K in a Quantum Design
measuring system.

Fig.~1 shows the resistivity of the oxygen-isotope exchanged films of (a)
LCMO; (b) NSMO over a temperature range 100-300 K. It
is striking that, in all the cases, the $^{18}$O samples have
lower metal-insulator crossover temperatures and much sharper
resistivity drop. The Curie temperature $T_{C}$ normally coincides with a
temperature where $d\ln\rho/dT$ exhibits a maximum. We find that the
oxygen isotope shift of $T_{C}$ is 15.0(6) K for LCMO, and 18.0(6)
K for NSMO, in excellent agreement with the
results for the bulk samples \cite{Zhao99}.

\begin{figure}[htb]
\ForceWidth{6.6cm}
	\centerline{\BoxedEPSF{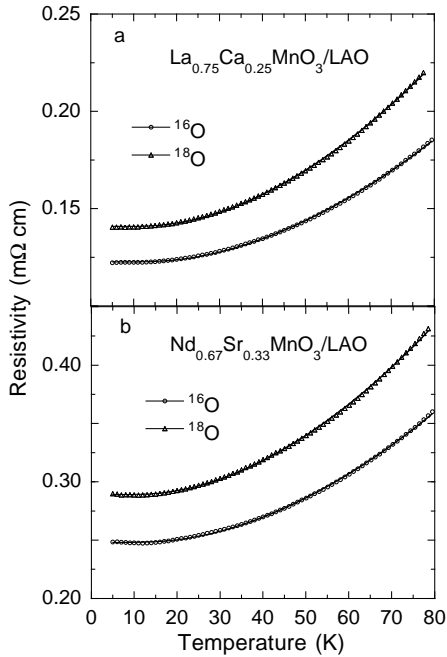}}
	\vspace{0.3cm}
	\caption[~]{The low-temperature resistivity of the oxygen-isotope
exchanged
	films of (a) La$_{0.75}$Ca$_{0.25}$MnO$_{3}$; (b)
Nd$_{0.7}$Sr$_{0.3}$MnO$_{3}$. The solid lines are fitted curves by
Eq.~1.}
	\protect\label{Fig.2}
\end{figure}
In Fig.~2 we plot the low-temperature resistivity of the oxygen-isotope
exchanged films of (a) LCMO; (b) NSMO. In both cases, the
residual resistivity $\rho_{0}$ for the $^{18}$O samples is larger than
for the $^{16}$O samples by about 15$\%$. We have repeated the van
der Pauw measurements at 5 K for several times with different
contact configurations. We checked that the uncertainty of the
difference in $\rho_{0}$ of the two isotope samples is
less than 3$\%$. We also found that the isotope effect is reversible
upon the isotope back-exchange. It is worth noting that the $\rho_{0}$
for the $^{16}$O NSMO film is nearly the same as for a single crystalline
sample \cite{Lee}. This indicates that the electrical transport
properties observed in our films are intrinsic.

The low temperature resistivity has been
explained by small polaron metallic conduction
\cite{Zhaopreprint}, which leads to a formula
\cite{Zhaopreprint,Lang}:
\begin{equation}\label{eq}
\rho(T) = \rho_{0} + B\omega_{s}/\sinh^{2}(\hbar\omega_{s}/2k_{B}T),
\end{equation}
where $\omega_{s}$ is the frequency of the softest optical
mode (about 100 K \cite{Zhaopreprint}), and $B$ is a constant, being
proportional to $m_{p}/n$ (here $m_{p}$ is
the effective mass of polarons, and $n$ is the mobile carrier
concentration). The residual resistivity
$\rho_{0}$ $\propto$ $m_{p}/n\tau_{0}$ \cite{Austin}, where
$\hbar/\tau_{0}$ is the zero temperature
scattering rate which is associated with the random potential
produced by randomly distributed trivalent and divalent cations
\cite{Pick}. The magnitude of $\hbar/\tau_{0}$ has been found to
be 18(3) meV from
optical data of the compounds Nd$_{0.7}$Sr$_{0.3}$MnO$_{3}$
\cite{Lee}, La$_{0.7}$Sr$_{0.3}$MnO$_{3}$ and
La$_{0.7}$Ca$_{0.3}$MnO$_{3}$ \cite{Simpson}.
Since the Drude weight ($\propto$ $n/m_{p}$) for
Nd$_{0.7}$Sr$_{0.3}$MnO$_{3}$ \cite{Lee} is about a factor of 3
smaller than for La$_{0.7}$Ca$_{0.3}$MnO$_{3}$ \cite{Simpson}, nearly the
same $\hbar/\tau_{0}$ observed for both compounds implies that
the $\hbar/\tau_{0}$ is nearly independent of $m_{p}/n$, and that
both $\rho_{0}$ and $B$ are proportional to $m_{p}/n$.

The best fit to the data shown in Fig. 2 indeed demonstrates that,
in both cases, $\rho_{0}$ increases by 15(3)$\%$, and $B$ by 18(3).
The results indicate that there is a substantial oxygen isotope effect on
$m_{p}/n$. Since we have shown that \cite{ZhaoPRL,Zhao99} the total
hole concentrations for two isotope samples should be the same, the
observed isotope effect on $m_{p}/n$ implies that $m_{p}$ depends
strongly on the isotope mass. This strongly suggests that the carriers in the
ferromagnetic state are indeed small polarons, in
agreement with the theoretical model \cite{Alex}. We also find that both
$\rho_{0}$ and
$B$ for NSMO are about twice as large as
those for LCMO, implying that the
effective mass of carriers for NSMO is larger
by a factor of 2.4 than for LCMO. This is
consistent with specific heat data; the electronic specific heat coefficient
$\gamma$ $\sim$ 8 mJ/moleK$^{2}$ for La$_{0.8}$Ca$_{0.2}$MnO$_{3}$
\cite{Hamilton}, while $\gamma$ $\sim$ 20 mJ/moleK$^{2}$ for
Nd$_{0.67}$Sr$_{0.33}$MnO$_{3}$ \cite{Gordon}.

In the paramagnetic state, the dominant conduction mechanism is
thermally activated hopping of adiabatic small polarons. The mobility
is given by \cite{Emin}
\begin{equation}
\mu = \frac{ed^{2}}{h}\frac{\hbar\omega_{o}}{k_{B}T}\exp (-
E_{a}/k_{B}T),
\end{equation}
where $d$ is the site to site hopping distance, which is equal to
$a/\sqrt{2}$ since the carriers in manganites mainly reside on the oxygen
sites \cite{Ju}, $\omega_{o}$ is the characteristic frequency of the optical
phonons, and $E_{a}$ is given by \cite{Mott,Emin,Lang1} $E_{a} =
(\eta E_{p}/2)f(T) - t$,
where $E_{p}$ is the polaron binding energy, $t$ is the ``bare'' hopping
integral, $\eta$ is a constant ($\eta$ = 1 for Holstein polarons
\cite{Emin} and
$\eta$ $\sim$ 0.2-0.4 for Fr\"ohlich polarons \cite{Alexcond}),
and $f(T) = [\tanh (\hbar\omega_{o}/4k_{B}T)]/
(\hbar\omega_{o}/4k_{B}T)$ \cite{Mott}.

If one considers that
small polarons can be bound into localized pairs (bipolarons)
\cite{Alex} or to the
impurity centers \cite{Austin}, one finds \cite{Alex,Austin}
that the mobile carrier density
$n = 2(2\pi m_{p}k_{B}T/h^{2})^{3/2}\exp(- \Delta /2k_{B}T)$
when $T$$<$$W_{p}/k_{B}$ and
for a parabolic band, where $W_{p}$ is the
polaron bandwidth ($m_{p} = 6\hbar^{2}/a^{2}W_{p}$), $\Delta$ is
the bipolaron binding energy \cite{Alex} or twice
the gap between localized impurity levels and the bottom of the polaron
band \cite{Austin}. In fact, the above $n(T)$ expression is the same
as that for semiconductors when the chemical potential is pinned to
the mobility edge of the bipolarons or to the impurity levels.
For a single-type (hole or electron) carrier system, $\Delta$ = 2$E_{s}$,
where
$E_{s}$ is the activation energy deduced from thermopower
data. Then in the temperature
region: $T_{C}$$<$$T$$<$$W_{p}/k_{B}$, the resistivity $\rho = 1/ne\mu$ is
given by
\begin{equation}
\rho = \frac{A}{\sqrt{T}}\exp(E_{\rho}/k_{B}T),
\end{equation}
where $A = (ah/e^{2}\sqrt{k_{B}})(1.05W_{p})^{1.5}/\hbar\omega_{o}$,
and $E_{\rho} = E_{a} + \Delta /2$. The quantity $A$
should strongly depend on the
isotope mass because $W_{p}
= W_{o}\exp(-\Gamma E_{p}/\hbar\omega_{o}) =
W_{o}\exp(-g^{2})$. Here $W_{o}$ is the bare bandwidth ($W_{o}$ = 12$t$),
$\Gamma$ is
a constant, which is less than 0.4 for both Holstein and Fr\"ohlich
polarons when the coupling constant $\lambda = 2E_{p}/W_{o}$$<$0.5
and $\hbar\omega_{o}/t$$<$1 \cite{Alex99}. For
$T$$>$$>$$W_{p}/k_{B}$, the exponential prefactor in Eq.~3 will be
proportional
to $T/\omega_{o}$ \cite{Jaime}. In this case, the exponent of the
isotope effect on
the prefactor is about -0.5, in sharp contrast to a large and positive
exponent predicted by  Eq.~3.

In Fig. 3, we show the resistivity in the paramagnetic state for the
$^{16}$O and $^{18}$O films of (a) LCMO; (b) NSMO. It is clear that
the resistivity strongly depends on the oxygen isotope
mass. The data can be well fitted by Eq.~3 with the parameters: $A$ =
17.3/12.7 m$\Omega$cm K$^{0.5}$, $E_{\rho}/k_{B}$ = 844.3/997.2 K for the
LCMO $^{16}$O/$^{18}$O film; $A$ =
22.2/16.2 m$\Omega$cm K$^{0.5}$, $E_{\rho}/k_{B}$ = 925/1082 K for the
NSMO $^{16}$O/$^{18}$O film.  For LCMO, the parameter $A$ decreases by
35(5)$\%$, and $E_{\rho}$ increases by 13.2(5) meV. For NSMO, $A$ decreases by
37(8)$\%$, and $E_{\rho}$ increases by 13.5(9) meV. The observed large
oxygen isotope effect on the parameter $A$ is consistent with Eq.~3.
It is worth noting that since the activation energy is large, one can also
fit the data by $\rho
(T) = AT^{\beta}\exp(E_{\rho}/k_{B}T)$ with $\beta$ varying from -2 to
2.

We can use the values of the parameter $A$ for the $^{16}$O films to calculate
the polaron bandwidth $W_{p}$ according to the
relation: $A = (ah/e^{2}\sqrt{k_{B}})(1.05W_{p})^{1.5}/\hbar\omega_{o}$.
By taking $\hbar\omega_{o}$ = 74
meV, typical for oxides \cite{Alex99}, we yield $W_{p}$ = 49 meV for
the LCMO $^{16}$O film, and 58 meV for the NSMO $^{16}$O film. The polaron
bandwidth is greatly reduced compared with the bare bandwith $W_{o}$,
which can be estimated to about 4.9 eV from the relation:
$W_{o} = 6\hbar^{2}/a^{2}m_{b}$ and $m_{b}$ = 0.61$m_{e}$ \cite{Zhaopreprint}.
Then we can determine $g^{2}$ to be 4.61/4.44 for the LCMO/NSMO $^{16}$O film
using $g^{2} = \ln (W_{o}/W_{p})$. If $\hbar\omega_{o}$ decreases by 5.7$\%$
upon replacing $^{16}$O by
$^{18}$O, then our calculation shows that the parameter $A$ decreases
by 35$\%$ for LCMO, and by 33$\%$ for the
NSMO, in good agreement with the measured values:
35(5)$\%$ for LCMO and 37(8)$\%$ for NSMO.

It is important to clarify whether the small polarons are bound into
localized pairs (bipolarons) or just to the impurity centers since
this can place an essential experimental constraint on various CMR
theories. If the small polarons are bound to the impurity centers,
there will be no isotope effect on $\Delta$ \cite{Austin}. In this
case, the isotope
effect on $E_{\rho}$ only arises from the isotope shift of $E_{a}$,
as seen from the relation: $E_{\rho} = E_{a} + \Delta /2$. Since
$E_{a} = (\eta E_{p}/2)f(T) - W_{o}/12$, only the quantity
$f(T) = [\tanh (\hbar\omega_{o}/4k_{B}T)]/
(\hbar\omega_{o}/4k_{B}T)$ may depend on the isotope mass if the
temperature is not so high compared with $\hbar\omega_{o}/k_{B}$. If we
take $\hbar\omega_{o}$ = 74 meV, we have $f(T)$ = 0.855 and $\delta f(T)$ =
0.0134
for T = 300 K, where $\delta$
means a change due to the oxygen isotope substitution. Using
the relation: $(\eta E_{p}/2)f(T) = W_{o}/12 + E_{\rho} - \Delta
/2$, and $\Delta /2$ = $E_{s}$ = 15 meV for La$_{0.75}$Ca$_{0.25}$MnO$_{3}$
\cite{Hundley}, we yield $\eta
E_{p}/2$ = 0.54 eV. Then $\delta E_{\rho} = \delta E_{a}
= (\eta E_{p}/2)\delta f(T)$ = 7.2 meV, which is about half the value
observed.
\begin{figure}[htb]
    \ForceWidth{6.6cm}
	\centerline{\BoxedEPSF{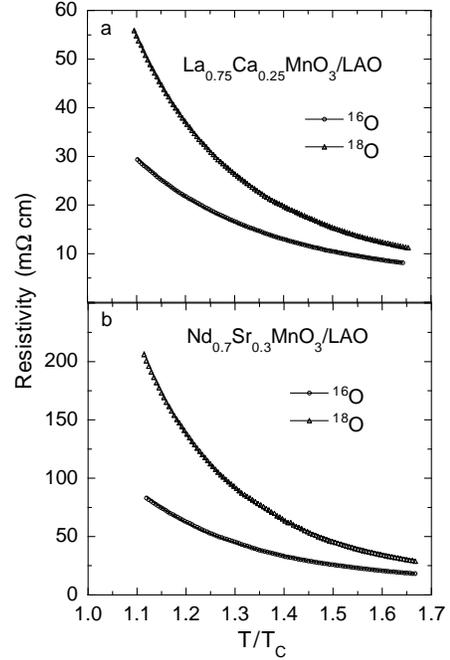}}
	\vspace{0.3cm}
	\caption[~]{The resistivity in the paramagnetic state for the
$^{16}$O and $^{18}$O films of (a) La$_{0.75}$Ca$_{0.25}$MnO$_{3}$; (b)
Nd$_{0.7}$Sr$_{0.3}$MnO$_{3}$. The solid lines are the fitted curves
by Eq.~3. Note that the fit is not so perfect for the NSMO $^{18}$O
film since the data are not very smooth around T = 1.4$T_{C}$.}
	\protect\label{Fig.3}
\end{figure}

On the other hand, one can quantitatively explain the
isotope dependence of $E_{\rho}$ if small polarons form localized
bipolarons. In this scenario, $\delta\Delta = - \delta W_{p}$
\cite{Alexcond}, so $\delta\Delta  = 0.057g^{2}W_{p}$.
From the deduced values for $g^{2}$ and $W_{p}$ above, we
calculate that $\delta\Delta$ = 12.9 meV for LCMO, and 14.7 meV for
NSMO. Then  $\delta E_{\rho} = \delta E_{a} + \delta\Delta /2$ = 13.6
meV for LCMO, and 14.5 meV for NSMO. The calculated values are in excellent
agreement with the observed values: 13.2(5) meV for LCMO and 13.5(9) meV
for NSMO. In addition, if the
small polarons are bound into small bipolarons, there
will be a maximum in optical conductivity at the energy $\nu_{b} =
2g^{2}\hbar\omega_{o}+ W_{p} + \Delta $ \cite{Alexcond}.
Substituting $g^{2}$ = 4.61, $\hbar\omega_{o}$ = 74 meV, $W_{p}$ = 49
meV and $\Delta$ = 30 meV for the LCMO $^{16}$O film, we
obtain $\nu_{b}$ = 0.76 eV, in excellent agreement
with experimental results (i.e., $\nu_{b}$ $\simeq$ 0.70 eV for $x$ = 0.3,
and 0.80 eV for $x$ = 0.1) \cite{Jung}.

In summary, the oxygen isotope effects observed in high-quality
epitaxial thin films of
La$_{0.75}$Ca$_{0.25}$MnO$_{3}$ and Nd$_{0.7}$Sr$_{0.3}$MnO$_{3}$ can
be quantitatively explained by a scenario \cite{Alex} where the small polarons
form localized bound pairs (bipolarons) in the paramagnetic state
while the carriers in the ferromagnetic states are small polarons due
to the competing exchange interaction
of polaronic carriers with localized spins. The electron-phonon coupling
strength $g^{2}$
in the paramagnetic state is about twice of that in
the low temperature ferromagnetic state. This is possibly because a
strong screening effect by mobile charge carriers reduces the
electron-phonon interaction in the ferromagnetic state. The
present results provide quantitative constraints on the basic physics
of manganites.

{\bf Acknowlegement}: We would like to thank A. S. Alexandrov and A. M.
Bratkovsky for useful discussions. We also thank X. M. Zhang for
helping with resistivity measurements. The work was supported by the
NSF MRSEC at the University of Maryland and Swiss National Science
Foundation.
~\\
~\\
$*$ Present address: Laboratoire CRISMAT-ISMRA,14050 CAEN Cedex, France.
\bibliographystyle{prsty}

\begin{thebibliography}{10}
\bibitem{Von} R.~von~Helmolt, J.~Wecker, B.~Holzapfel, L.~Schultz, and
K.~Samwer, Phys. Rev. Lett. \textbf{71}, 2331 (1993); S.~Jin, T. H. Tiefel,
M. McCormack, R. A. Fastnacht, R.
Ramesh, and L. H. Chen, Science \textbf{264}, 413 (1994).
\bibitem{Art}A. P. Ramirez, J. Phys.:Condens. Matter, \textbf{9}, 8171 (1997).
\bibitem{Zener} C. Zener, Phys. Rev. \textbf{82}, 403 (1951); P. W.
Anderson, and H. Hasegawa,  Phys. Rev.
\textbf{100}, 675 (1955).
\bibitem{Millis1} A. J. Millis, P. B. Littlewood, and B. I. Shraiman,
Phys. Rev. Lett. \textbf{74}, 5144 (1995);A. J. Millis, B. I.
Shraiman, and R. M\"uller, Phys. Rev. Lett. \textbf{77}, 175 (1996).
\bibitem{Roder} H. R\"oder, J. Zang,  and A. R. Bishop, Phys. Rev.
Lett. \textbf{76},
1356 (1996).
\bibitem{Moreo} A. Moreo, S. Yunoki, and E. Dagotto, Science
\textbf{283}, 2034 (1994).
\bibitem{Alex} A. S. Alexandrov and A. M. Bratkovsky, Phys. Rev.
Lett. \textbf{82},
141 (1999).
\bibitem{Jaime} M. Jaime, M. B. Salamon, M. Rubinstein, R. E. Treece,
J. S. Horwitz, and D. B. Chrisey, Phys. Rev. B \textbf{54},
11914 (1996).
\bibitem{ZhaoNature}G. M. Zhao, K. Conder, H. Keller, and
K. A. M\"uller, Nature (London) \textbf{381}, 676 (1996).
\bibitem{ZhaoPRL} G. M. Zhao, M. B. Hunt and H.~ Keller,
Phys. Rev. Lett. \textbf{78}, 955 (1997).
\bibitem{ZhaoJPCM}G. M. Zhao, K. Ghosh, and R. L. Greene, J.
Phys.: Condens. Matter, \textbf{10}, L737 (1998).
\bibitem{Billinge} S. J. L. Billinge, R. G. DiFrancesco, G. H. Kwei,
J. J. Neumeier, and J. D. Thompson, Phys. Rev. Lett.
\textbf{77}, 715 (1996).
\bibitem{Teresa} J. M. De Teresa, M. R. Ibarra, P. A. Algarabel, C.
Ritter, C. Marquina, J. Blasco, J. Garcia, A. del Moral, and Z.
Arnold, Nature (London), \textbf{386}, 256 (1997).
\bibitem{Booth} C. H. Booth, F. Bridges, G. H. Kwei, J. M. Lawrence,
A. L. Cornelius, and J. J. Neumeier, Phys. Rev. Lett.
\textbf{80}, 853 (1998).
\bibitem{Prellier}W. Prellier, M. Rajeswari, T. Venkatesan and R. L.
Greene, App. Phys. Lett. \textbf{75},
1146 (1999).
\bibitem{Zhao99} G. M. Zhao, K. Conder, H. Keller, and
K. A. M\"uller, Phys. Rev.
B \textbf{60}, 11 914 (1999).
\bibitem{Lee}H. J. Lee, J. H. Jung, Y. S. Lee, J. S. Ahn, T. W. Noh, K. H.
Kim, andS. W. Cheong, Phys. Rev. B \textbf{60}, 5251 (1999).
\bibitem{Zhaopreprint}G. M.  Zhao, V. Smolyaninova, W. Prellier, and H.
Keller, cond-mat/9912037
\bibitem{Lang} I. G. Lang and Yu. A. Firsov, Sov. Phys. -JETP
\textbf{16}, 1301 (1963); V. N. Bogomolov, E. K. Kudinov, and Yu. A.
Firsov, Sov. Phys. - Solid State \textbf{9}, 2502 (1968).
\bibitem{Austin}I. G. Austin and N. F. Mott, Adv. Phys. \textbf{18},
41 (1969)
\bibitem{Pick} W. E. Pickett and D. J. Singh, Phys. Rev. B
\textbf{55}, R8642 (1997); D. A. Papaconstantopoulos and W. E.
Pickett, Phys. Rev. B \textbf{57}, 12751 (1998)
\bibitem{Simpson} J. R. Simpson, H. D. Drew, V. N. Smolyninova, R. L.
Greene, M. C. Robson, A. Biswas, and M. Rajeswari, cond-mat/9908419.
\bibitem{Hamilton}J. J. Hamilton, E. L. Keatley, H. L. Ju, A. K.
Raychauduri, V. N. Smolyninova, and R. L. Greene, Phys. Rev. B
\textbf{54}, 14926 (1996)
\bibitem{Gordon} J. E. Gordon, R. A. Fisher, Y. X. Jia, N. E.
Phillips, S. F. Reklis, D. A. Wright, and A. Zettl, Phys.
Rev. B \textbf{59}, 127 (1999).
\bibitem{Emin} D. Emin and T. Holstein, Ann. Phys. (N.Y.), \textbf{53},
439 (1969).
\bibitem{Ju} H. L. Ju, H. C. Sohn, and K. M. Krishnan, Phys. Rev. Lett.
\textbf{79}, 3230 (1997).
\bibitem{Mott}N. F. Mott and E. A. Davis ~{\em Electronic Processes in
Non-crystalline Materials} (Clarendon Press, Oxford, 1979) p. 82
\bibitem{Lang1} I. G. Lang and Yu. A. Firsov, Sov. Phys. -JETP
\textbf{27}, 443 (1968)
\bibitem{Alexcond}A. S. Alexandrov and A. M. Bratkovsky,
J. Phys.: Condens. Matter, \textbf{11}, L531 (1999)
\bibitem{Alex99} A. S. Alexandrov and P. E. Kornilovitch, Phys. Rev.
Lett. \textbf{82},
807 (1999).
\bibitem{Hundley}M. F. Hundley and J. J. Neumeier, Phys. Rev.
B \textbf{55}, 11 511 (1997).
\bibitem{Jung}J. H. Jung,  K. H. Kim, T. W. Noh, E. J. Choi, and J. J.
Yu, Phys. Rev. B \textbf{57}, R11 043 (1998).



\end{thebibliography}

\end{document}